\documentclass[a4paper,10pt,3p,twocolumn]{elsarticle}

\usepackage{graphicx}
\usepackage{hyperref}
\usepackage{siunitx}
\usepackage[version=4]{mhchem}
\usepackage{caption}
\usepackage{subcaption}
\usepackage{amsmath}
\usepackage{amssymb}

\journal{Astroparticle Physics}
\begin{document}

\begin{frontmatter}

\title{Identification of Patterns in Cosmic-Ray Arrival Directions using Dynamic Graph Convolutional Neural Networks}

\author{T.~Bister}
\author{M.~Erdmann}
\ead{erdmann@physik.rwth-aachen.de}
\author{J.~Glombitza}
\author{N.~Langner}
\author{J.~Schulte}
\author{M.~Wirtz}

\address{RWTH Aachen University, III. Physikalisches Institut A, Otto-Blumenthal-Str., 52056 Aachen, Germany}

\begin{abstract}
We present a new approach for the identification of ultra-high energy cosmic rays from sources using dynamic graph convolutional neural networks. These networks are designed to handle sparsely arranged objects and to exploit their short- and long-range correlations. Our method searches for patterns in the arrival directions of cosmic rays, which are expected to result from coherent deflections in cosmic magnetic fields. The network discriminates astrophysical scenarios with source signatures from those with only isotropically distributed cosmic rays and allows for the identification of cosmic rays that belong to a deflection pattern. We use simulated astrophysical scenarios where the source density is the only free parameter to show how density limits can be derived. We apply this method to a public data set from the AGASA Observatory.
\end{abstract}

\begin{keyword}
ultra-high energy cosmic rays \sep sources \sep magnetic fields \sep neural networks 
\end{keyword}

\end{frontmatter}

\section{Introduction}
\label{sec:introduction}
The quest for sources of ultra-high energy cosmic rays has entered a new phase. On the one hand, the experimental results of recent years have sharpened the boundary conditions for this search; on the other hand, current developments in data analysis technologies are opening up new perspectives. 

The measurements relevant to source searches include first the sharp drop in the energy spectrum above $50$~EeV~\cite{Abraham:2008ru,Abraham:2010mj,AbuZayyad:2012ru,ThePierreAuger:2015rha}. Second, measurements of the slant depth of air showers in the atmosphere indicate a mixed composition of protons and heavier nuclei~\cite{Aab:2014kda,Aab:2014aea,Aab:2017cgk}. Both these observations increase the probability of astrophysical scenarios in which the sources are some megaparsecs away and accelerate nuclei to a maximum rigidity (=energy/charge)~\cite{Aab:2016zth}. The third relevant observation is the discovery of a large-scale dipole signal above 8~EeV energy which shows - with a significance of more than five standard deviations - that the arrival directions of cosmic rays are not entirely isotropic~\cite{Aab:2017tyv,Aab:2018mmi}. Also, there is an indication of an intermediate-scale anisotropy: the correlation of particles with energies above $39$~EeV with starburst galaxies exhibits an increasing significance, currently at the 4.5$\sigma$ level~\cite{Aab:2018chp,Caccianiga:2019xyz}.

For nuclei with charge $Z$, deflections in cosmic magnetic fields of several degrees per charge unit are estimated~\cite{Stanev:1996qj,Harari:2000az,Harari:2002dy,Golup:2009cv,Giacinti:2010dk,Jiang_2010,Golup:2011tv,Giacinti:2011uj}, with the strongest deflection expected in the magnetic field of our Galaxy~\cite{Dolag:2003ra,Paoletti:2010rx,Durrer:2013pga,Hackstein:2016pwa,Bray:2018ipq}. The different rigidities of the nuclei may lead to patterns in the arrival directions which can be used to distinguish cosmic rays of a single source or of a concatenated source cluster from isotropically arriving particles~\cite{Abreu:2011md,Aab:2014dha}.

In the field of data analysis, new technologies around machine learning are the driving force of much lot of the progress. Deep-learning methods~\cite{Goodfellow:2016:DL:3086952} in particular have gained considerable attention in astroparticle physics \cite{Erdmann:2017str,Erdmann:2018kuh,Shilon:2018xlp,Delaquis:2018zqi,Fuhrer:2018euq,Guillen:2018uex,Erdmann:2019nie,Anton:2019wmi,Bezyazeekov:2019jbe,Kronmueller:2019jzh,Kalashev:2019skq}. 

Recently, we introduced a fit method which contracts patterns in the arrival directions induced by the Galactic magnetic field to determine most probable extragalactic source directions~\cite{Erdmann:2018cvz}. In this fit, several thousand parameters are determined simultaneously (source directions, particle charges), which was realized by using the backpropagation method developed for neural network training~\cite{Tensorflow:2016}.

In this paper we present a new method for source identification that directly uses pattern recognition in cosmic-ray arrival directions. A similar approach using convolutional neural networks was developed in parallel \cite{Kalashev:2019skq}. Here, we use the concept of so-called dynamic graph convolutional neural networks~\cite{DGCNN:2018} which have already been successfully adapted and applied to challenges in particle physics~\cite{Qu:2019gqs,Toptagger:2019}. In contrast to our fit method~\cite{Erdmann:2018cvz}, here we do not use an explicit model of the Galactic magnetic field, but instead train the network with typical deflection patterns resulting from these field models~\cite{Pshirkov:2011um,Jansson:2012pc,Pshirkov:2013wka,Jansson:2012rt,Beck:2014pma,Han_2018}. 

The scope of our method is to search for the existence of patterns in cosmic-ray arrival directions: if such patterns are detected by the network, we classify which individual particles originate from a common source and which can be attributed to isotropic arrival directions. For this classification we analyze variables autonomously formed by the network. 

If no pattern is found, we derive a lower limit on the density of cosmic-ray sources using astrophysical simulations with the source density as their only free parameter. We exploit public data of the AGASA Observatory to demonstrate the application to real cosmic-ray measurements.

This work is structured as follows. First, we present astrophysical benchmark simulations used for training and performance estimation as well as the public dataset of the AGASA Observatory. We then introduce the functionality of dynamic graph convolutional neural networks. We describe the search for the existence of anisotropic patterns, the subsequent identification of cosmic rays and the determination of source density limits. Finally, we apply our methods to the data measured by the AGASA Observatory. We complete the work with our conclusions.

\section{Cosmic rays: simulations and measurements}
\label{sec:data}
In this section we first present the simulated data we use for training the network and evaluating its performance. 

For an in-depth understanding of the network, we start with simplified simulations in which a single source generates a few cosmic rays that exhibit typical signal patterns in the arrival directions due to the deflection in cosmic magnetic fields. As background we further add isotropically arriving cosmic rays.

Later, in an advanced simulation, we use an astrophysical scenario that depends only on one free parameter, the density of sources. Here, all cosmic rays originate from these sources, with only a few closer sources contributing several cosmic rays which form a pattern in the arrival distributions. This results in a few signal patterns distributed across the whole sky, each with particles from a common source, while most particles appear to have isotropic arrival directions.

Finally, we present the AGASA Observatory and its public data set of ultra-high energy cosmic rays. 

\subsection{Simplified cosmic-ray scenario with a single source}
\label{sec:astro_scenario_1}
In order to keep the results unambigous and clean, this first simulation setup is simplified yet fairly in agreement with the measurements described in section~\ref{sec:introduction}.

During the training process we generate up to 300.000 simulation sets on-the-fly with the following general setup. The predominant part of all $N_\textrm{CRs}$ events, called background, is positioned isotropically on the sky. 
We then add a certain fraction $f_\textrm{S}=N_\textrm{S}/N_\textrm{CRs}$ of $N_\textrm{S}$ signal cosmic rays which mimic a pattern arising from deflections in the Galactic magnetic field (GMF). Current models of the GMF consist of a superposition of a \textit{regular} field part which deflects particles coherently and a \textit{turbulent} field part which results in blurring and widening of the incoming particle beam. The magnitude of the magnetic field influence depends on the cosmic-ray rigidity \begin{equation}
    R=\frac{E}{Z}\;\; , 
\end{equation}
the ratio of its energy and charge. To prevent overtraining of the network which could learn specific patterns in fixed directions of the sky we do not use one fixed field parametrization. Instead, we use a random value for the orientation of the signal pattern and adopt the field magnitude of both regular and turbulent fields from~\cite{Jansson:2012pc, Jansson:2012rt} in the following way: we account for the turbulent field by a Fisher~\cite{Fisher1953} distribution with a rigidity-dependent width of 
\begin{equation}
    \label{eq:turb_fixed}
    \sigma_{\textrm{turb}}(R) = \frac{T}{R~/~\textrm{EV}}~\textrm{rad}\, ,
\end{equation}
where we choose $T$ position-dependently based on typical scattering obtained in \cite{Jansson:2012pc, Jansson:2012rt}. During the training we fix $T=2.8$ which corresponds to half the maximum of the expected turbulent deflection in ~\cite{Jansson:2012pc, Jansson:2012rt} and more than double the mean value. It was tested that training with a fixed value $T$ for all training sets is beneficial for the network performance even when evaluating with varying values of~$T$.

The coherent deflection angle $\delta_{\textrm{coh}}$ is described by
\begin{equation}
    \label{eq:coherent_deflection}
    \delta_{\textrm{coh}}(R) = \frac{D}{R~/~\textrm{EV}}~\textrm{rad} \, ,
\end{equation}
with a varying deflection power $D$ which is determined position-dependently based on typical deflections from~\cite{Jansson:2012pc}. For the training it is taken from a Gaussian distribution based on deflections again from~\cite{Jansson:2012pc}. Additionally, a lower cut on $D$ ensures that the coherent deflection mostly remains larger than the turbulent one.

The cosmic-ray energies follow the parameterized measured energy spectrum given in~\cite{Fenu:2017} which contains a broken power law with a smooth transition function above the ankle. The energy threshold of 40~EeV is oriented on the data used for the recent first indication of intermediate-scale anisotropy in cosmic-ray arrival directions~\cite{Aab:2018chp}. At these high energies typical analyses of modern observatories contain around $\mathcal{O}(1000)$ events which our simulation accounts for with a total event number of $N_{\textrm{CRs}}=1000$.

\begin{figure}[htbp]
\begin{center}
\includegraphics[width=0.49\textwidth]{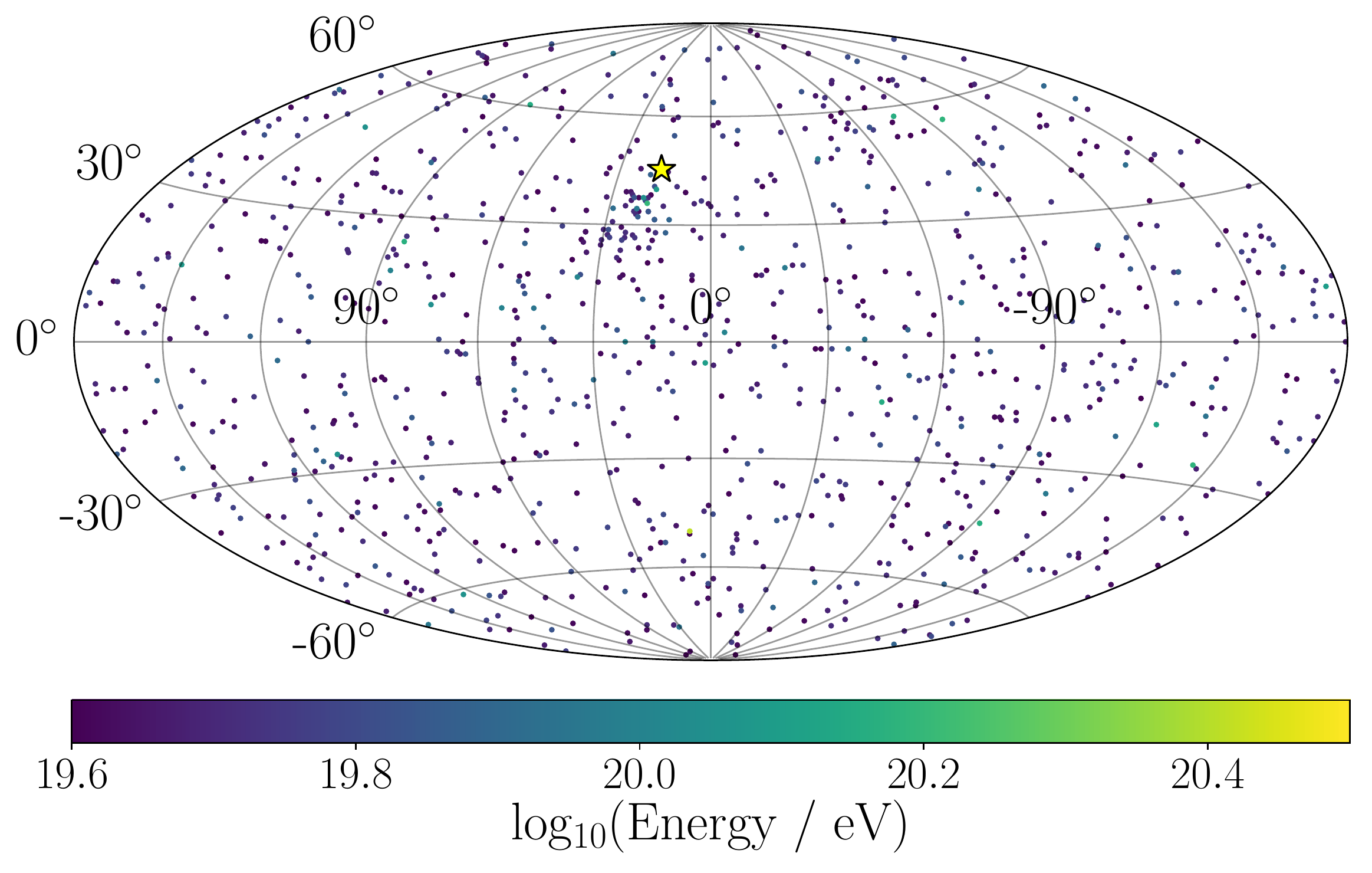}
\caption{Example of simulated arrival directions used for network training: the pattern induced by a magnetic field  with a coherent deflection power $D=7.2$ is situated at Galactic longitude $l\approx16^\circ$ and Galactic latitude $b\approx45^\circ$. The source of this signal pattern is indicated by the star symbol. It has a signal fraction of $f_\textrm{S}=5.5\%$, thus $N_\textrm{S}=55$ signal cosmic rays are contained in this arrival pattern. The remaining $945$ events are isotropically distributed.}
\label{fig:arrival_directions}
\end{center}
\end{figure}

A pure helium composition is chosen for the network training to enable the learning of comprehensive energy orderings. Also, having a mono-atomic composition ensures that the network does not artificially learn ratios between elemental fractions which could differ between simulations and data. Nevertheless, using helium instead of protons accounts for the measured heavier component of cosmic rays at the highest energies. An example simulation is shown in Fig.~\ref{fig:arrival_directions}.

To evaluate the network performance we also simulate a mixed composition scenario of the same kind using $15\%$ hydrogen, $45\%$ helium and $40\%$ carbon, nitrogen and oxygen (in equal parts) nuclei. 

\subsection{A general astrophysical scenario originating from a source population}
\label{sec:astrophysical-scenario}
In the previous section we introduced a simple astrophysical simulation consisting of two parts, an isotropic background and an additional magnetic-field induced arrival pattern with an arbitrarily fixed number of signal cosmic rays. In the following we extend this simplified approach to constrain actual physics quantities using the network.

A common assumption is to consider a uniformly distributed source population with source density $\rho_\textrm{S} = N_{\textrm{sources}}~/~V$, thus exhibiting an average number of $N_{\textrm{sources}}$ source candidates in any given spatial volume $V$. Each of the sources accelerates nuclei of nuclear fractions $f_A$ with a simple power-law energy spectrum of spectral index $\gamma$ and maximum rigidity cut-off $R_{\text{cut}}$:
\begin{equation}
\frac{\text{d}N_A}{\text{d}E} \propto f_A \cdot E^{-\gamma} \cdot f_{\text{cut}} (E, Z_A 
R_{\text{cut}})
\end{equation}
These nuclei are emitted isotropically and they propagate through the universe before arriving on Earth. More details about the source modeling, e.g. the cut-off function $f_{\text{cut}}$, can be found in \citep{Aab:2016zth}. 

Under the conditions that each source $i$ emits the same number of nuclei with the same energy spectrum and the extragalactic magnetic field is weak compared to the Galactic one, the arriving particle flux $f$ per source above an energy threshold at the edge of our Galaxy depends on only two aspects. 

The first one is the geometrical effect in a three-dimensional propagation scenario where the flux $f_i$ of a source $i$ at distance $d_i$ decreases with $f_i~\propto~d_i^{-2}$ while the source number per distance shell increases by $N_{\textrm{sources}}\propto d_i^{2}$.
Thus, the geometrical effect alone would result in a constant flux per distance bin. 

The second important effect is interactions during cosmic-ray propagation, e.g. with photon fields, where particles lose energy and the composition changes. Additionally, nuclear decay and cosmological effects can attenuate particles. When combined, these processes cause a suppression of particles from faraway sources. Thus, overall this scenario will again result in some prominent sources and a practically isotropic background of farther sources of which only very few particles arrive on Earth.

The parameters describing the sources within this general universe setup have been evaluated before, using a combined fit of energy spectrum and composition of one-dimensional \textit{CRPropa3}~\citep{Batista2016} simulations to cosmic-ray data measured by the Pierre Auger Observatory~\citep{Aab:2016zth}. In summary, the determined source parameters were: a spectral index of $\gamma = 0.87$, a rigidity cut-off at $\log_{10}(R_{\text{cut}}/\text{V}) = 18.62$, and a composition containing 88 percent nitrogen and 12 percent silicon (for details see~\citep{Aab:2016zth}). When arriving on Earth, the composition was altered due to propagation effects.

We now combine these one-dimensional propagation results with the described three-dimensional simulation setup including both geometrical and interaction effects. The only free parameter is the source density $\rho_\textrm{S}$ which regulates the number and strength of the patterns induced by nearby sources. For a small source density there are only few and therefore very prominent nearby sources; this results in rather obvious anisotropic arrival patterns within an isotropic background. For high source densities we therefore expect predominantly isotropic arrival directions.

Having these extragalactic arrival directions we can treat deflections by the GMF analogously as in section~\ref{sec:astro_scenario_1}. Therefore, the magnitude of the coherent and of the turbulent deflection respectively again follow~\cite{Jansson:2012pc, Jansson:2012rt}.
Different to the first simulation this time cosmic rays originate from more than one source and thus a related deflection model has to be applied on the sphere as a whole. 
This is ensured by using a HEALPix~map~\citep{healpy1, healpy2} of the direction of the coherent deflection in~\cite{Jansson:2012pc, Jansson:2012rt}.

\begin{figure}[htbp]
\begin{center}
\includegraphics[width=0.49\textwidth]{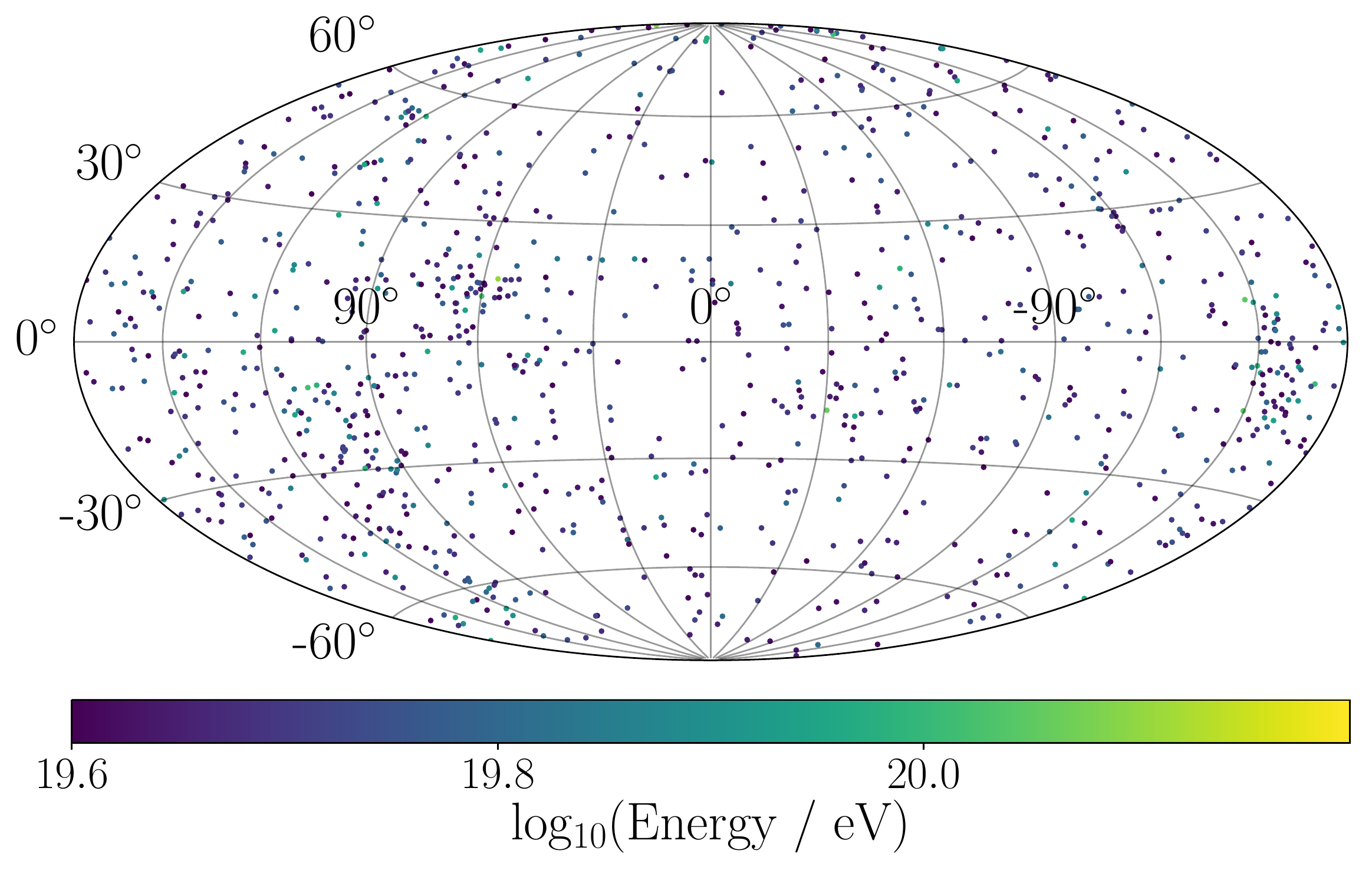}
\caption{Example of simulated arrival directions at a source density of $10^{-3}/\mathrm{Mpc}^3$. For this rather small source density an anisotropic distribution of cosmic rays is obvious. Note that the energy spectrum and composition of arriving cosmic rays is in accordance with measurements~\citep{Aab:2016zth}.}
\label{fig:arrival_directions2}
\end{center}
\end{figure}

During training, $\sigma_\mathrm{turb}(R)$ is again chosen independently of the cosmic-ray direction as described in Eq.~(\ref{eq:turb_fixed}) keeping $T$ fixed. Additionally, a random individual rotation is applied to the direction and coherent deflection maps, resulting in different deflections in every generated sample. An example arrival directions distribution is shown in Fig.~\ref{fig:arrival_directions2}.

\subsection{Dataset of the AGASA Observatory}
\label{sec:AGASA-data}
The Akeno Giant Air Shower Array (AGASA) consisted of $111$ surface detectors deployed over an area of about $100$~km$^2$ and additionally $27$ muon detectors. AGASA was situated at $138^\circ\; 30^\prime$ E and $35^\circ\; 47^\prime$ N. This location determines the shape of the exposure of the observatory which we calculate according to \cite{Sommers:2000us}.
The observatory was put into operation in 1990 \cite{Chiba:1991nf,Ohoka:1996ww} and was dismantled in 2007~\cite{ICRR:2007}.

The public event list for cosmic rays with energies above $E=40$~EeV and zenith angles up to $\theta=45^\circ$ contains $57$ events in total \cite{Takeda_1999,Hayashida:2000zr}.

\section{Graph Convolutional Neural Networks}
\label{sec:DGCNN}

In the following we will first present an overview of the deep learning techniques necessary for this analysis and then show the architecture of the dynamic graph convolutional neural network used.

\subsection{Convolutional neural networks}
Traditional \textit{convolutional neural networks}~\cite{convolutions} are most successful e.g. in computer vision and pattern recognition. The dense, discrete and regular collection of image pixels allows for scanning the image with pixelated filters of a constant extent. Usually, $K$ filters (kernels), covering only $k_n$ neighboring pixels, i.e. a small part of the overall image, are used. The convolution of a single filter with index $a$ applied to pixel $i$ of a single channel image reads:
\begin{equation}
\label{eq:convolution}
    f_i^a = \sum_{j=1}^{k_n} \theta_j^a x_{i_{j}}
\end{equation}
with the result $f_i^a$ stored in a \textit{feature map}, which in turn is scanned by further filters after adding a bias and applying an activation function. In addition to Cartesian grids, spherical grids have also been used in convolutional networks \cite{Perraudin:2018rbt,Kalashev:2019skq} based on the healpix pixelization \cite{healpy2}.

Usually, filters for deeper layers receive information from more distant pixels due to the increasing receptive field of view, so that in a figurative sense short-range correlations can be examined in the first layers and long-range correlations in the deeper layers. All $K$ filters indexed $a$ with their respective parameters are trained on the basis of a task formulated in the objective function.

\subsection{Graph convolutional networks}
A problem for applications of CNNs in astroparticle physics is that the arrival directions of cosmic rays are continuously distributed. If one wants to position them artificially on a grid corresponding to the experimental directional resolution, the pixel occupancy is extremely sparse on the one hand and, on the other hand, in a few pixels several particles may be found which requires an algorithm to aggregate the information. Both aspects are rather unfavorable for the actual application as this leads to loss of information or major computational costs. Furthermore, most convolution algorithms are designed for Euclidean manifolds and therefore cannot handle the symmetry of spherical data. 

The concept of \textit{graph convolutional networks} (GCN)~\cite{Hamilton2017, kipf2016semi} solves the unnatural requirement of particles placed on a regular grid. 
In GCNs, each particle can be treated individually with its arrival direction. All particles together form a point cloud and, when using a specific neighborhood assignment, a graph. Here, the exact alignment and position of the particles has to be considered in the convolutional operation and in the structure of the graph. In this astroparticle-physics application, the graph is constructed in a spherical shape as the cosmic rays arrive almost uniformly from all directions onto Earth. The particles in immediate proximity to each other in the coordinate space are then regarded as the environment for a convolution.
In contrast to standard convolutional networks, the explicit cosmic ray can still be identified in the deeper network layer allowing for node classification and high interpretability.

\subsection{Dynamic graph convolutional neural networks}

\begin{figure}[htbp]
\begin{center}
\includegraphics[width=0.42\textwidth]{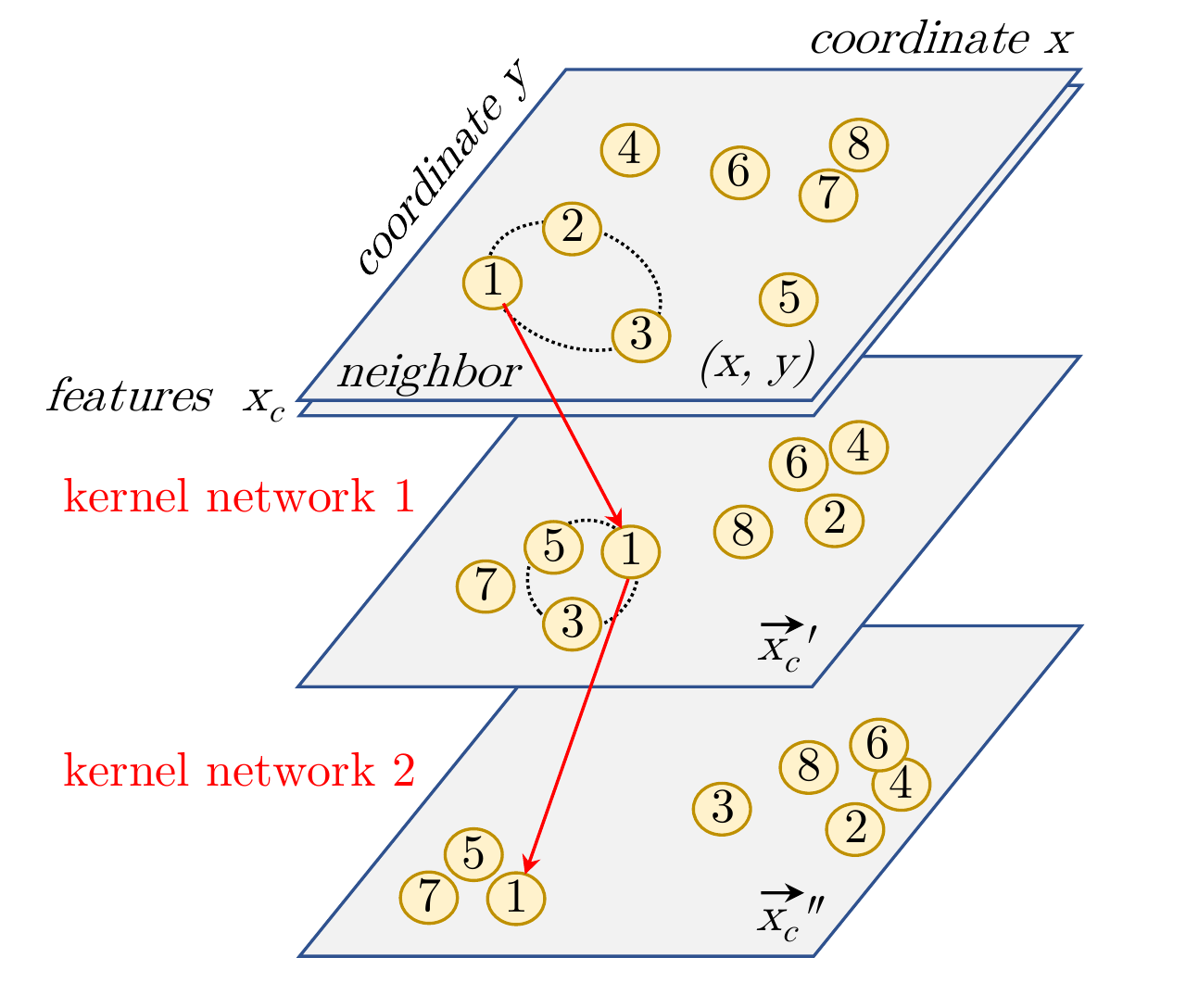}
\caption{Principle of a dynamic graph convolutional neural network with the example of $8$ cosmic rays with arrival coordinates $(x,y)$ and properties $\vec{x}_c$ (features). The convolution operation is performed by a neural network on the coordinates and features, using the $2$ nearest neighbors of each particle. The result of the operation is placed in a new high-dimensional space, thereby changing the neighborhood of the cosmic rays. In this way, the arrival patterns of the cosmic rays, even if they are distributed over the entire sphere, can be jointly characterized immediately in the following network layer.}
\label{fig:dgcnn}
\end{center}
\end{figure}

The special feature of \textit{dynamic graph} convolutional neural networks (DGCNNs) is that in each layer the original graph is projected onto another graph with arbitrary dimension in coordinate space as illustrated in Fig.~\ref{fig:dgcnn}. Each particle can still be followed through the network, however, its nearest neighbors have changed. The high-dimensional coordinates which for each transition define the nearest neighbors and thus the graph are derived from the properties (features) $\vec{x}_{c}$ of the particles, e.g. arrival direction, energy, shower depth, etc.

For the convolution using the kernel, the $k_n$ nearest neighbors of a particle $i$ are considered. Particle $i$ has $M$-dimensional properties $\vec{x}_{i,c}$ which will be related to the properties of the $k_n$ neighboring particles. The convolutional operation is implemented following~\cite{DGCNN:2018} using a neural network which depends on all values of $\vec{x}_{i,c}$, and the differences $\vec{x}_{i,c}-\vec{x}_{i_j,c}$ between the $M$ properties of the particle $i$ and those of its neighbor $i_j$:
\begin{align}
    h_\Theta(\vec{x}_{i,c}, \vec{x}_{i_j,c} ) = \mathrm{NN}_\Theta(\vec{x}_{i,c}, \vec{x}_{i,c}-\vec{x}_{i_j,c}),
    \label{eq:conv}
\end{align}
where $\Theta$ are trainable parameters of the kernel network $\mathrm{NN}_\Theta$.
The same kernel network is then applied to all $k_n$ neighbors. The final result of the convolution for particle $i$ can be obtained by an aggregation over the neighborhood, e.g. by calculating the average value:
\begin{align}
    \vec{x}'_{i,c} = \cfrac{1}{k_{n}} \sum_{j=1}^{k_n}  h_\Theta( \vec{x}_{i,c}, \vec{x}_{i_j,c}).
    \label{eq:edgeconv}
\end{align}
The output dimension of the kernel network then defines the new dimension of the coordinate space for the resulting new graph consisting of the $N_\mathrm{CRs}$ cosmic rays, illustrated in~Fig.~\ref{fig:dgcnn}. The original $N$-dimensional coordinate space $(x,y)$ is extended, resulting in a new graph of $K$ feature dimensions, denoted by $\vec{x}^\prime_c$ in the middle plane of Fig.~\ref{fig:dgcnn}. This is the decisive difference to a classic graph convolutional network: the clustered particles are not necessarily next to each other in their original spatial coordinates. Thus, new neighborhoods resulting from the similar properties of the particles are investigated in the forthcoming convolutional layer.

Also, the following convolutional layer is realized by a network (cf. Fig.~\ref{fig:dgcnn} lower level). As the rearranged particles inherit new neighbors, both short- and long-range correlations are already analyzed at this step. If a binary classifier is trained using the objective function, the parameters of the two neighboring networks will be adjusted such that two separate groups of particles become visible in the high-dimensional space denoted by $\vec{x}^{\prime\prime}_c$ in the figure.

\subsection{Network architecture}
\begin{figure}[htbp]
\begin{center}
\includegraphics[width=0.48\textwidth]{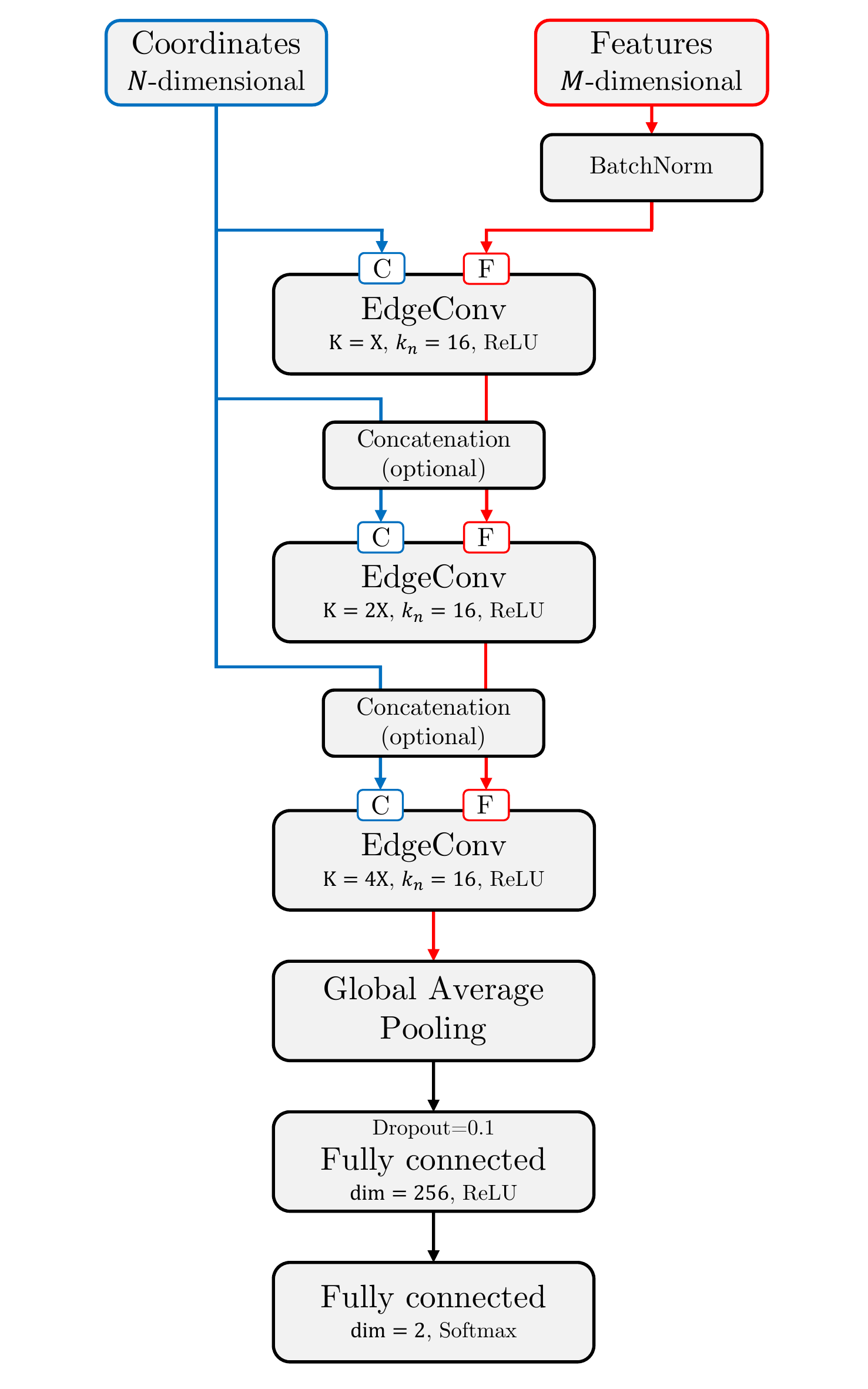}
\caption{Architecture of the applied neural network. The $k_n=16$ nearest neighbors of each particle are determined from the coordinate space $C$. The convolutional operations on the features $F$ are performed three times in \textit{EdgeConv} blocks. Optionally, neighbored particles are fixed through all calculations of the network or varied by concatenation of the spatial coordinates with the features obtained with the convolutional operations.}
\label{fig:architecture}
\end{center}
\end{figure}

Our implementation\footnote[1]{code available at \url{https://git.rwth-aachen.de/niklas.langner/edgeconv_keras}} of the network architecture is inspired by~\cite{Qu:2019gqs}. In this reference, Eq.~(\ref{eq:edgeconv}) is referred to as \textit{EdgeConv} layer~\cite{DGCNN:2018}. In Fig.~\ref{fig:architecture}, we show at the top the input with the $N$-dimensional coordinate space and the $M$-dimensional feature space.

Instead of expressing the cosmic-ray arrival direction by zenith and azimuth, we use the three-dimensional normalized vector $(x,y,z)$ of the arrival direction to avoid discontinuities in spherical coordinates. As input features we use the arrival direction together with the energy of the cosmic ray, thus four dimensions $(x,y,z,E)$. 

While the normalized directional vector $(x,y,z)$ has reasonable numerical values by definition, the strong variation in the values of the cosmic-ray energies is reduced prior to the network input. By using a transformation of the energy
$E'= E_0 / E^3$ with $E_0 = 175,600$~EeV$^3$, the energy distribution becomes more uniform with most values at $\mathcal{O}(1)$/EeV$^2$.

As the neighborhood we consider the $k_n=16$ nearest particles in coordinate space or in the combined coordinate and feature space, respectively. When varying this number we found no improvement for the tasks studied in this work. 

In the first EdgeConv layer, the $k_n$ nearest neighbors are calculated using the initial coordinates of the cosmic rays (input $C$ in Fig.~\ref{fig:architecture}). The convolutions are applied to their features (input $F$ in Fig.~\ref{fig:architecture}). Each EdgeConv layer performs one convolution as described in Eq.~(\ref{eq:edgeconv}) using a kernel network with three fully connected layers of the same dimension $K$. Batch normalization with momentum of $0.9$ and $\epsilon=10^{-5}$ as well as ReLU activation are applied after each layer. Finally, the average over all neighbors is taken to obtain a permutational invariant layer. The dimension $K$ of the convolution kernels is adjusted according to the specific task studied below. These successive transformations can be understood as a continuous and non-linear filter applied to each cosmic-ray neighborhood.

We use this network either as a graph convolutional neural network (GCN) by limiting the coordinates given to all subsequent EdgeConv layers to the cosmic-ray arrival directions, or alternatively as a dynamic graph convolutional neural network (DGCNN) by extending the coordinates with the feature space obtained in the previous EdgeConv layer. Only in the latter case do we allow new neighborhoods to be formed. Depending on the task to be solved, we find advantages for either of the network architectures.

After the convolutions, global average pooling is used to gather the global information contained in all cosmic rays. Using two fully-connected layers and a final softmax activation, a binary output is achieved which we call:
\begin{equation}
    \textrm{signal}: x_\mathrm{sig} \qquad \textrm{isotropy}: (1-x_\mathrm{sig})
\end{equation}
Thus, this final output of the network can be interpreted as a probability $x_\mathrm{sig}$ that the corresponding set of cosmic rays contains an arrival pattern which originated from a source.

\section{Searching for cosmic-ray sources}
\label{sec:patterns}
    
We evaluate the performance of the network in three challenges. In the first challenge, the goal is to recognize whether a signal pattern of a source is present within the arrival directions and energies of $1000$ cosmic rays. In an extension, the second challenge is to classify each particle as either belonging to a source or to an isotropic contribution. In the third task, the network should recognize whether there are signal patterns of several sources distributed across the sphere.

\subsection{Expected sensitivity for the signal pattern of a single source}
\label{sec:sensitivity}

In the first task we use the GCN (Fig.~\ref{fig:architecture}) with the arrival directions of the cosmic rays as fixed coordinates in all convolutional layers. This is meaningful as there is at most one coherent signal pattern on the sphere of arrival directions. An experiment with the DGCNN exhibited no improvement because there was no need to merge signal patterns scattered across the sphere.

We use $K=16$ convolution kernels in the first EdgeConv layer and increase this number in the following layers as shown in Fig.~\ref{fig:architecture}. After the global average pooling, a fully-connected layer with $256$ nodes is used. These hyperparameters were optimized by a scan. The categorical cross entropy was used as the objective function, and the Adam optimizer \cite{Kingma:2014vow} was used for training. The learning rate was scheduled to drop from $10^{-3}$ to $10^{-5}$ during the first two thirds of the training, following the falling range of the cosine function (between $0$ and $\pi$). The batch size was set to $60$ using $30$ sets of $1000$ arrival directions with a signal pattern of $55$ helium nuclei from a source and $30$ sets of $1000$ isotropically distributed arrival directions. The comparatively high signal fraction during training proved to be advantageous, since the fully trained network is then able to detect anisotropic arrival patterns even with much smaller signal fractions. Training experiments on different signal fractions or with a mixed composition provided no improvement. 

The expected sensitivity of the network prediction was evaluated on $1000$ signal sets each with a total of $1000$ cosmic rays, of which a varying number $N_\textrm{S}$ originate from a source and $10^7$ sets with isotropic arrival directions.
We show both the evaluation with helium nuclei and the evaluation with a mixed composition. Keep in mind that, in both cases, the network is trained with helium nuclei only. 

The final value $x_\mathrm{sig}$ of the softmax function for each of the signal sets was compared to the values of the isotropic sets. We define the $p$-value as the relative number of isotropic sets whose function value $x_\mathrm{sig}$ exceeds that of the signal set. 

\begin{figure}[htbp]
\begin{center}
\includegraphics[width=0.49\textwidth]{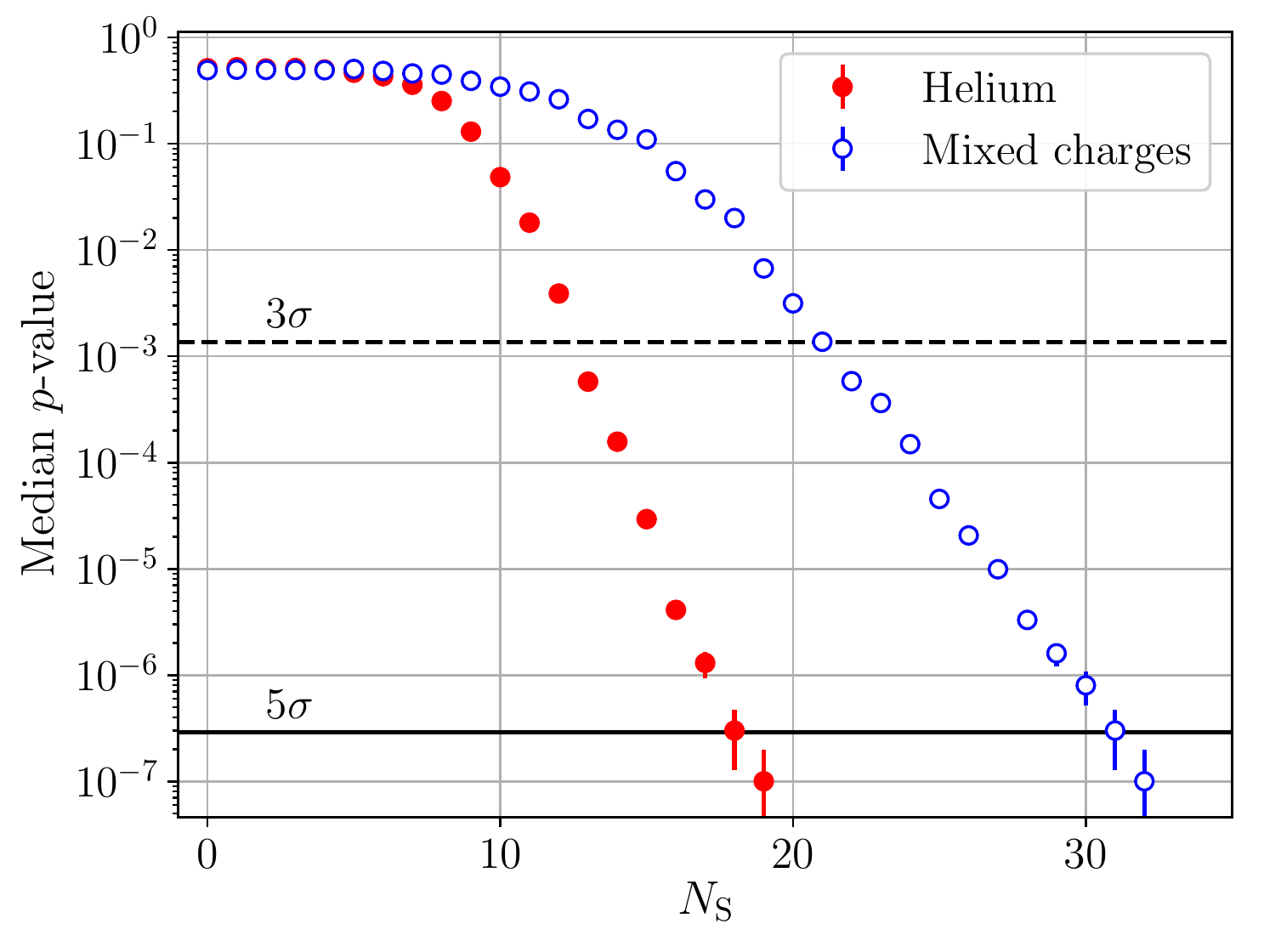}
\caption{Probability that isotropic cosmic-ray arrival directions result in a network response at least as high as a signal enriched simulation as a function of the number $N_\textrm{S}$ of source cosmic rays. The total number of events in one simulated dataset is 1000. Shown is the median $p$-value from $1000$ simulated evaluations for a helium composition (solid red symbols) and for a mixed composition (open blue symbols). The network was trained on signal scenarios with $N_S=55$ source nuclei and scenarios of solely isotropic arrival directions, both containing helium nuclei only.}
\label{fig:pvalue}
\end{center}
\end{figure}

In Fig.~\ref{fig:pvalue} we show the median of these $p$-values for all 1000 signal sets as a function of the number of simulated cosmic rays belonging to the signal pattern.

As expected, the mono-atomic case is easier as all cosmic rays of a pattern are ordered according to their energy and the charges have no additional influence on the deflection angle. Here, for $13$ source nuclei in $1000$ cosmic rays a signal pattern is detected with more than $3\sigma$ (standard deviations). The threshold of $5\sigma$ is reached with $18$ helium nuclei from one source. 

For a mixed composition, the minimum number of particles from one source increases to $21$ ($3\sigma$) or $31$ ($5\sigma$), respectively. These numbers are remarkable considering that the network never actually dealt with mixed charges during training. If the network were to exploit only the helium nuclei that it was trained on with a contribution of $45\%$ to the mixed-composition scenario, one would expect that $40$ cosmic rays would be needed for an effect of $5$ standard deviations. Thus, to achieve the same effect with fewer cosmic rays, the network also exploits information carried by the other nuclei.

\subsection{Identification of cosmic rays from a source}

In the second task, each individual cosmic ray will be examined to establish whether it belongs to a signal pattern and thus originates from a common source, or whether it is a particle of the isotropic arrival distribution. 

For this classification we analyze the autonomously formed variables of the network before the pooling layer as described in the following. Changing the network's output layer to perform a binary classification of each of the $1000$ cosmic rays yielded only slightly better results. For the benefit of an insight into the functioning of the original network, however, here we decide for cosmic-ray identification using this approach.

After the final EdgeConv layer with $64$ kernels ($K=4\times 16$, cf. Fig.~\ref{fig:architecture}), each cosmic ray is represented as a vector in a $64$-dimensional feature space resulting in a total dimensionality of (64 $\times$ $N_\textrm{CRs}$). The pooling layer takes the mean over all cosmic rays resulting in 64 output values. These encode the information about whether or not there is a signal pattern in the arrival directions of the cosmic rays, which is then processed by the subsequent fully connected layers to provide the final binary output. Instead, we now use the complete information of (64 $\times$ $N_\textrm{CRs}$) values before the pooling layer to identify cosmic rays originating from the source or from the isotropic arrival distribution.

\begin{figure}[t!]
\begin{center}
\begin{subfigure}{.99\textwidth}
\includegraphics[width=0.49\textwidth]{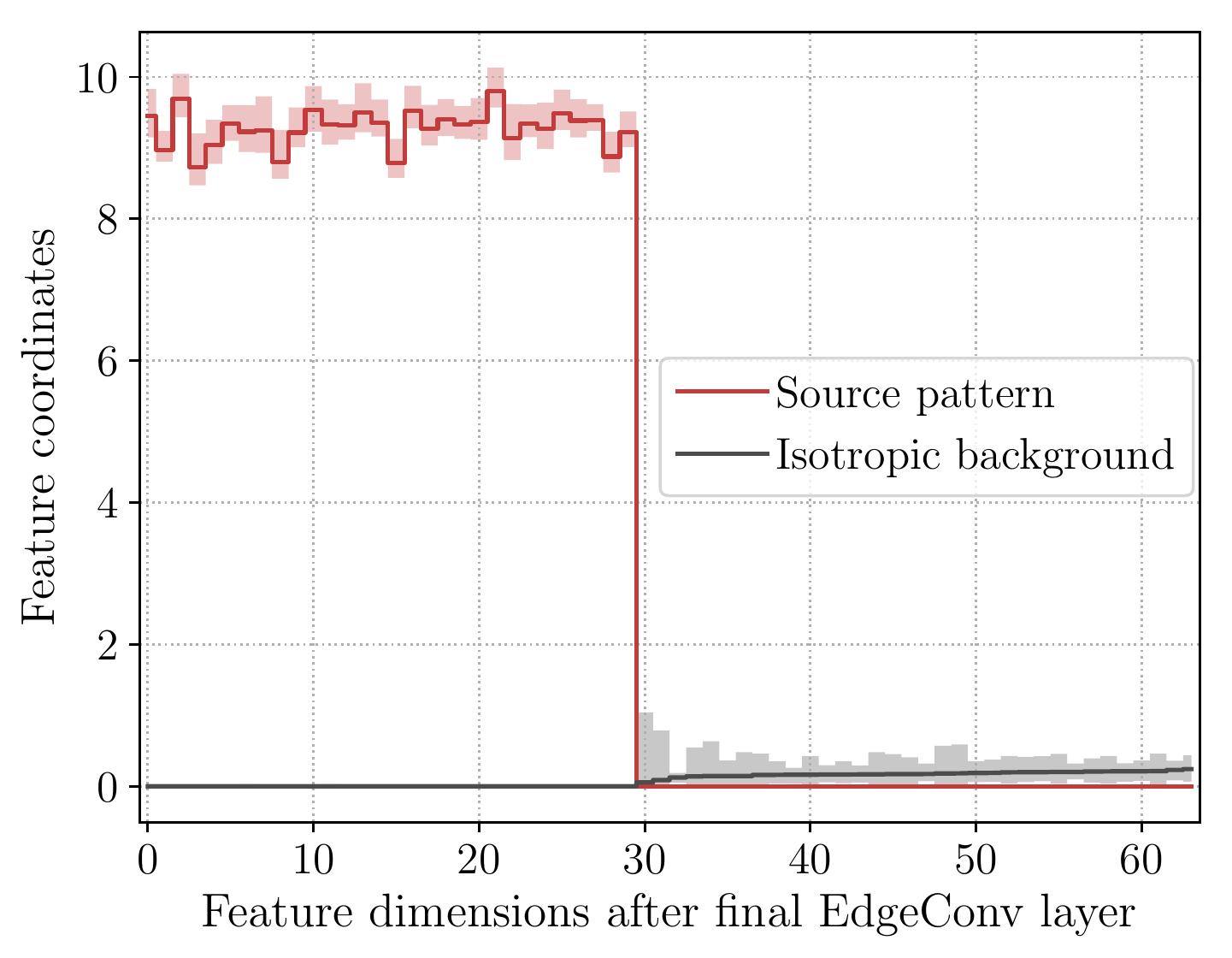}
\caption{}
\label{fig:median_dimension}
\end{subfigure}
\begin{subfigure}{.99\textwidth}
\includegraphics[width=0.49\textwidth]{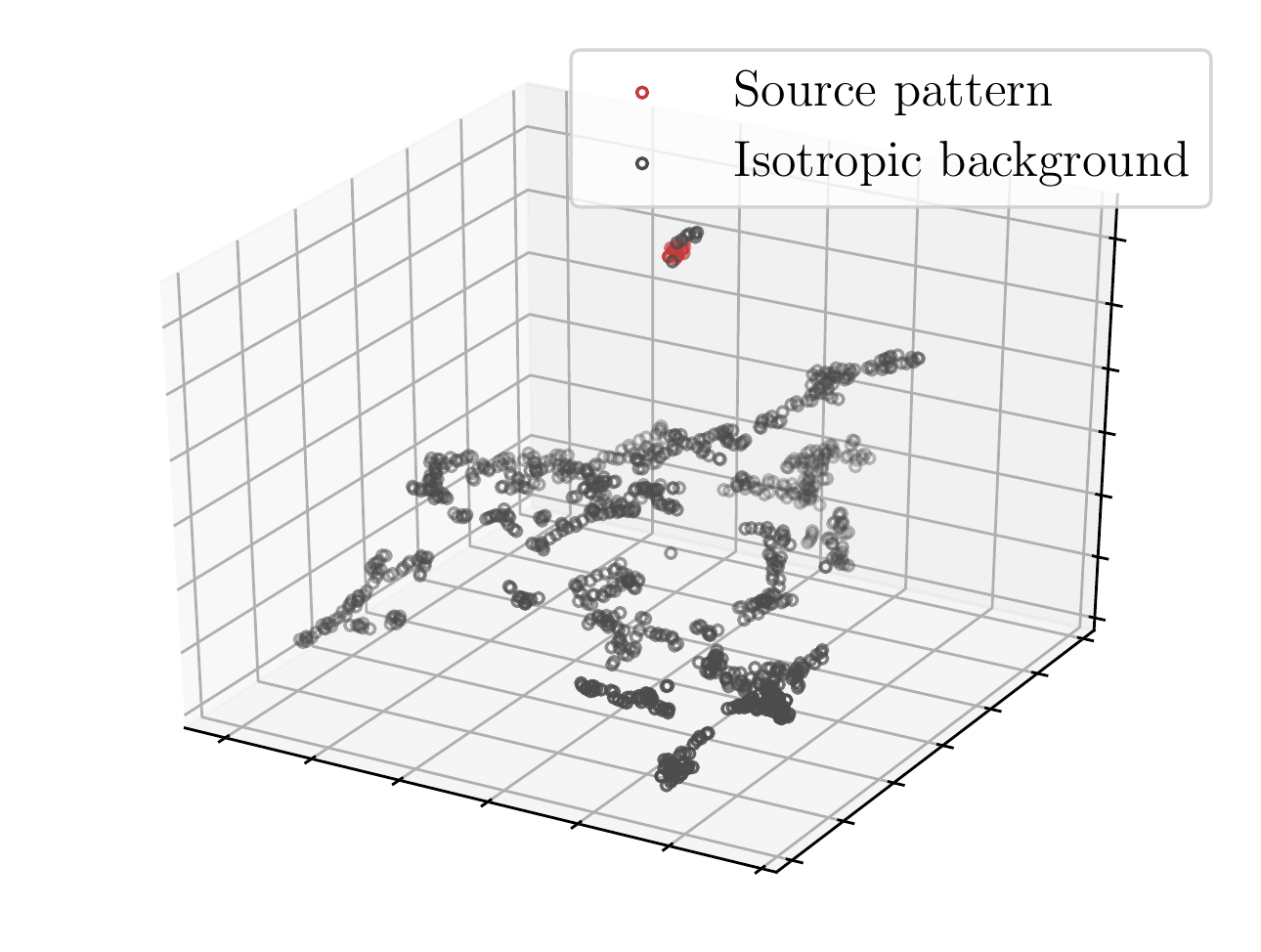}
\caption{}
\label{fig:tsne}
\end{subfigure}
\caption{Visualization of the network for separating signal cosmic rays (red) from isotropically arriving cosmic rays (gray). For a single simulated data set with $1000$ cosmic rays, $15$ of which originate from a common source the median values seen in (a) (solid lines) are calculated for each of the $64$ dimensions of the final EdgeConv layer. The shaded areas represent the 68\% quantile. (b) A three-dimensional representation of the final EdgeConv layer's output is created using the t-SNE technique~\cite{tsne}.}
\label{fig:dimension_identification}
\end{center}
\end{figure}

In Fig.~\ref{fig:median_dimension} we show the median values for each of the $64$ dimensions for a single set of $1000$ cosmic rays. The $15$ cosmic rays from the source are denoted by the dark-red marked distribution and the median values of isotropically arriving particles are shown as a distribution marked in gray. The order of the $64$ variables is arbitrary and shown here for visualization purposes in a way that the median value for the isotropic distribution increases. The colored areas around the median values denote the $68\%$ quantile of the values of all contributing cosmic rays.

For the particles from isotropic arrival distributions the median values are close to zero in all dimensions, and even invisible in the figure for dimensions $0$ to $29$ owing to their small values. For the particles of a signal pattern the information appears to be encoded in the dimensions $0$ to $29$ with median values around $9$. In the last $34$ dimensions, however, the median values of signal cosmic rays are close to zero. 
Thus, the network allocates approximately half of the available feature dimensions for cosmic rays originating from the source, while the other half is used for isotropic cosmic rays. This indicates that the network effectively performs clustering in the high-dimensional feature space which separates the signal cosmic rays from the isotropic ones.

For an additional visualization of this process we use the t-SNE algorithm. This algorithm embeds high-dimensional data in a reduced number of dimensions by keeping the implicit structure, i.e. neighborhoods and connections, in place. For further information refer to~\cite{tsne}. 

The three-dimensional representation of the 64-dimensional point cloud after the final Edge-Conv layer is shown in Fig.~\ref{fig:tsne}. One can see clear clustering of the cosmic rays from the source pattern, while some isotropic cosmic rays are also clustered near this region.

To distinguish between cosmic rays that most likely originated in a source and isotropic ones we make use of the network's clustering of source cosmic rays in the first dimensions of the feature space. Subsequently, for each cosmic ray we take the median of the first $27$ dimensions (preserving a buffer zone) of its vector in the feature space. We identify cosmic rays with a median value greater than $3$ as coming from a source and below $3$ as belonging to the isotropically arriving distribution. We have verified that this threshold works well even if the total number of cosmic rays varies.

\begin{figure}[htbp]
\begin{center}
\includegraphics[width=0.49\textwidth]{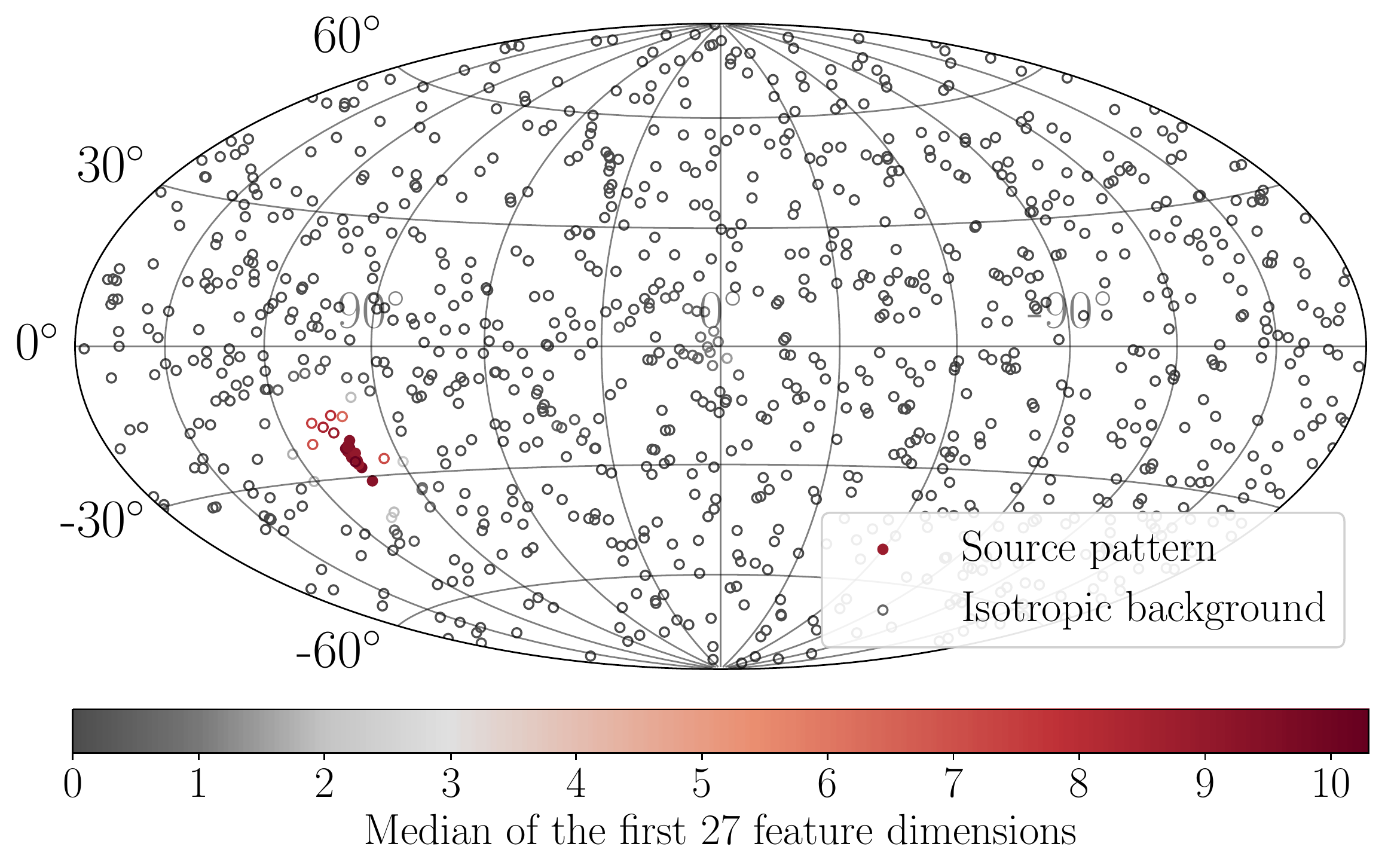}
\caption{Classified arrival directions: the color scale indicates the median of the first 27 feature dimensions. A cosmic ray is identified as signal with a median of the $27$ feature values greater than $3$ and is otherwise attributed to isotropy. The signal pattern (denoted by filled circles) is correctly identified.}
\label{fig:classification_skymap}
\end{center}
\end{figure}

To assess the quality of the identification, we calculate the efficiency of assigning a signal particle as belonging to the signal pattern by:
\begin{equation}
    \epsilon=\frac{\text{\# correctly identified signal cosmic rays}}{\text{\# signal cosmic rays}}
\end{equation}
We also calculate the purity with which a particle identified as signal actually belongs to the signal pattern of a source as:
\begin{equation}
    \rho=\frac{\text{\# correctly identified signal cosmic rays}}{\text{\# identified signal cosmic rays}}
\end{equation}

An example of classified arrival directions for a pure helium simulation is shown in Fig.~\ref{fig:classification_skymap}. The simulated underlying signal pattern containing 15 particles is identified and all signal particles were labeled correctly with an efficiency of 100$\%$. Some surrounding isotropic particles were incorrectly classified in this example with a median greater than $3$, which reduces the purity to 65.2$\%$.

\begin{figure}[t!]
\begin{center}
\begin{subfigure}{.99\textwidth}
\includegraphics[width=0.49\textwidth]{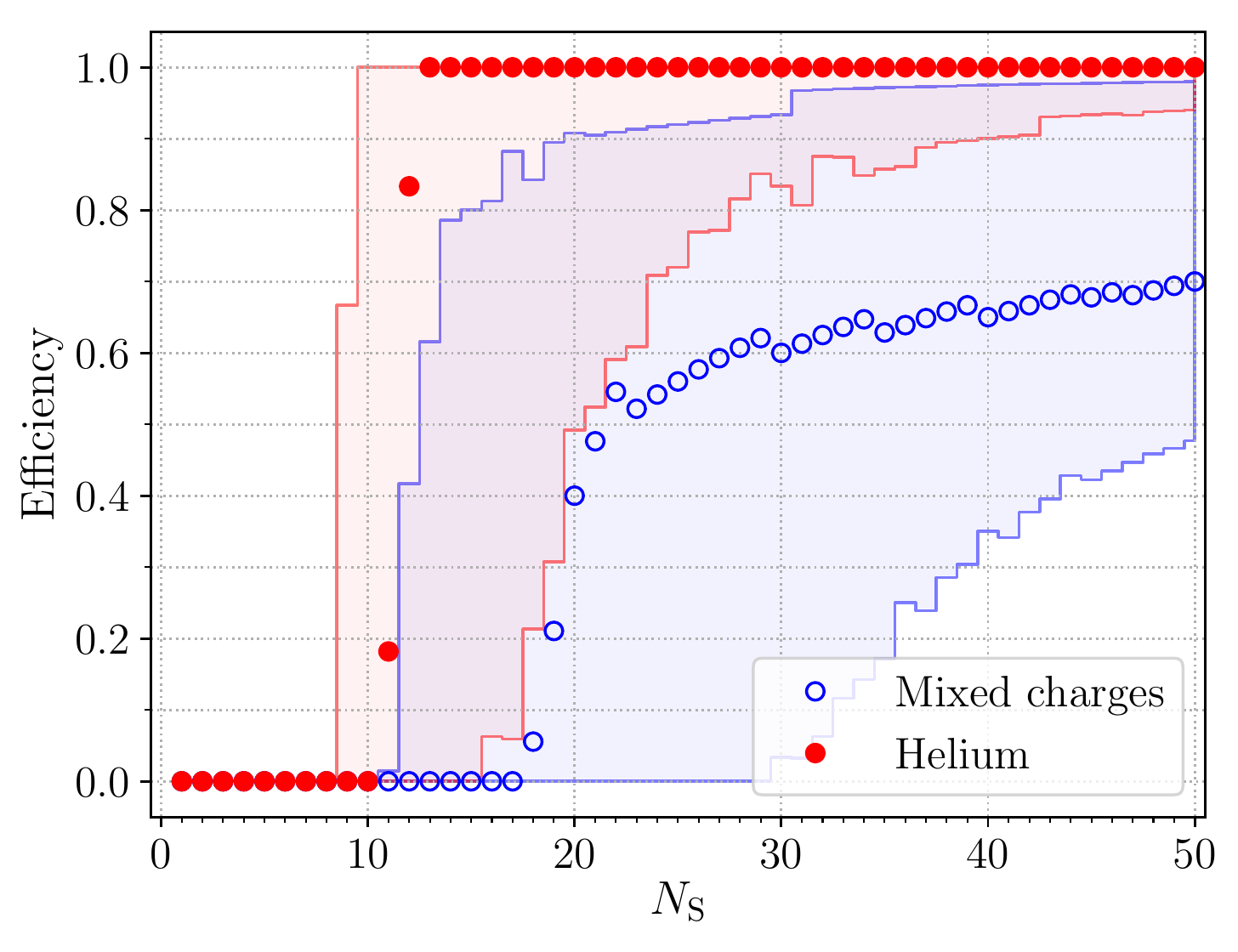}
\caption{}
\label{fig:efficiency1}
\end{subfigure}
\begin{subfigure}{.99\textwidth}
\includegraphics[width=0.49\textwidth]{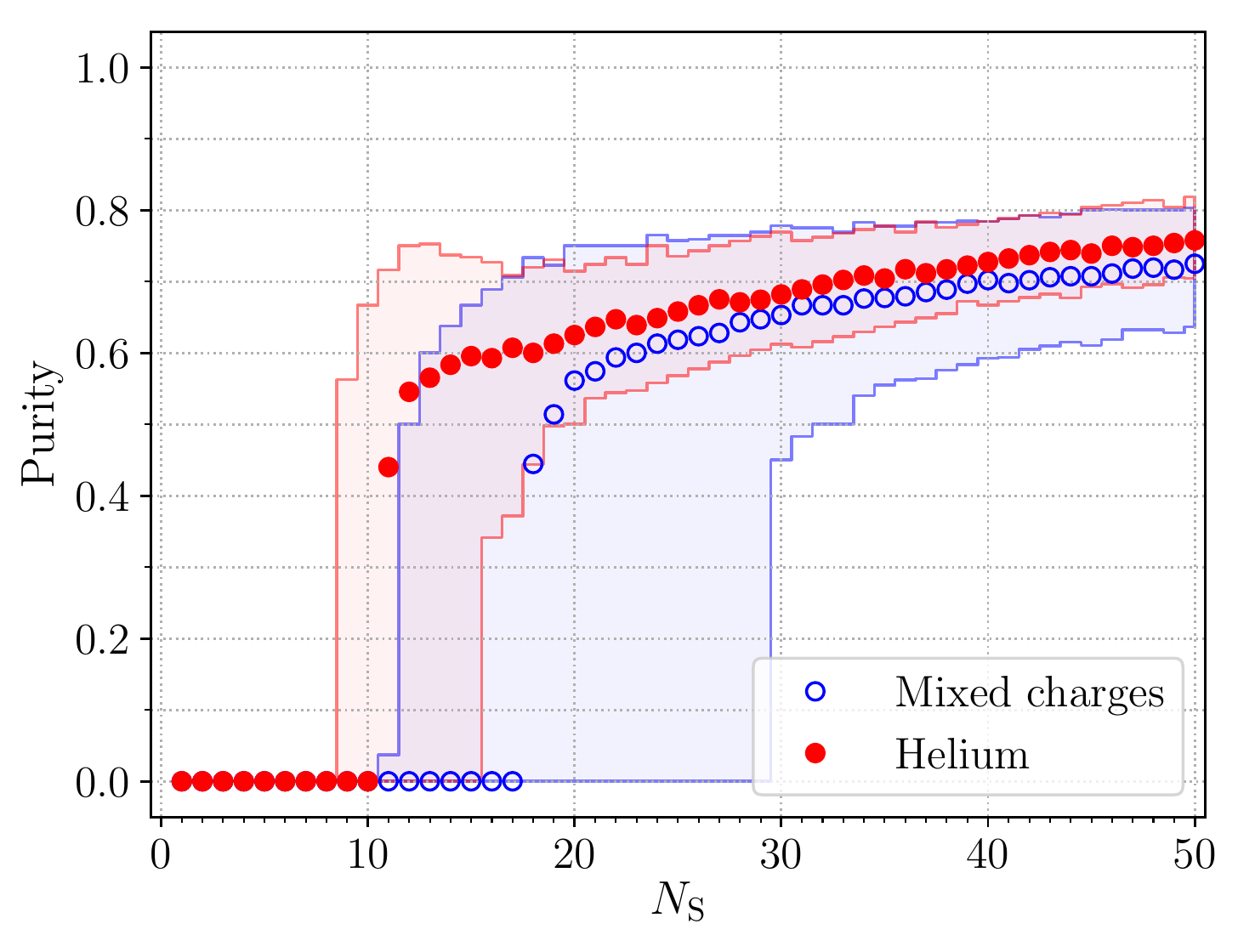}
\caption{}
\label{fig:purity1}
\end{subfigure}
\caption{Median efficiency (a) and purity (b) of 500 simulated sets as a function of the number of signal cosmic rays $N_\mathrm{S}$ coming from a single source. The remaining $(1000-N_\mathrm{S})$ cosmic rays are distributed isotropically. Both the pure helium (solid red) and the mixed composition (blue) are shown. Transparent bands represent the $68\%$ quantile.}
\label{fig:efficiency_purity}
\end{center}
\end{figure}

To assess whether or not this is a representative example we examine 500 sets with varying numbers of signal cosmic rays. In Fig.~\ref{fig:efficiency_purity} we show the efficiency and purity as a function of the number of signal particles of a source separately for the scenario with pure helium and for the mixed composition.

For the helium scenario, $18$ signal particles from one source were required to achieve a significance of $5$ standard deviations (cf. Fig.~\ref{fig:pvalue}). If the pattern is detected, in median all 18 cosmic rays are identified correctly with an efficiency of 100 percent as shown in Fig.~\ref{fig:efficiency1}. Nevertheless there are cases where fewer or even no cosmic rays with median feature dimension values above $3$ are detected resulting in cases with a low efficiency. This is shown by the shaded region representing the 68\% interval in Fig.~\ref{fig:efficiency1}. The median purity of 60\% shown in Fig.~\ref{fig:purity1} is still rather low at $N_\textrm{S}=18$ and exhibits a large spread. This means that, in addition to the (in median) $18$ correctly identified source particles, (in median) $12$ of the isotropically arriving particles are incorrectly identified as signals. In the mixed composition case, $31$ identified source particles for $5$ standard deviations have an efficiency and purity of around $60\%-70\%$.

\subsection{Search for multiple cosmic-ray sources}

In the third challenge, we again aim to identify signal patterns in the arrival directions and energies of cosmic rays. Due to the general astrophysical scenario used (see section~\ref{sec:astrophysical-scenario}), there may be signal patterns from several different sources, each contributing different numbers of cosmic rays.

For this analysis we use the dynamic graph convolutional network (DGCNN) with the extension of the coordinate space from the second network layer onward by considering the arrival directions of the particles as well as the features resulting from both the particles and their neighborhoods. Thus, the network has the possibility to change neighborhoods in deeper layers in such a way that neighborhood properties can be exploited for the respective challenge. Here, we find that the dynamic graph network performs better than the graph network with fixed neighborhoods. Since signal patterns can appear anywhere on the arrival sphere of the particles, the dynamic graph network can effectively analyze the signal patterns of different localizations on the sphere at an early stage by forming suitable neighborhoods. Optimized by scanning we decide to increase the kernel size from $K=64$ in the first EdgeConv layer to $K=256$ in the last one.

As training data we use simulated astrophysical scenarios with a source density of $10^{-3}$ Mpc$^{-3}$. At this rather small source density the arrival directions exhibit stronger anisotropies (cf. Fig.~\ref{fig:arrival_directions2}) which enables a stable training process. It was verified in a scan over the source density that $10^{-3}$ Mpc$^{-3}$ is a reasonable choice.

As isotropic comparison we also use the same scenarios, but randomize the arrival directions. In this way we ensure that the energy distribution of the cosmic rays remains unchanged.

We follow the previously outlined calculation of the $p$-value: first, we calculate the network prediction $x_\mathrm{sig}$ of $1000$ simulated astrophysical scenarios with signal patterns from different sources. We then determine the relative number of isotropic scenarios that have a larger value $x_\mathrm{sig}$ than each of the signal scenarios.

\begin{figure}[htbp]
\begin{center}
\includegraphics[width=0.49\textwidth]{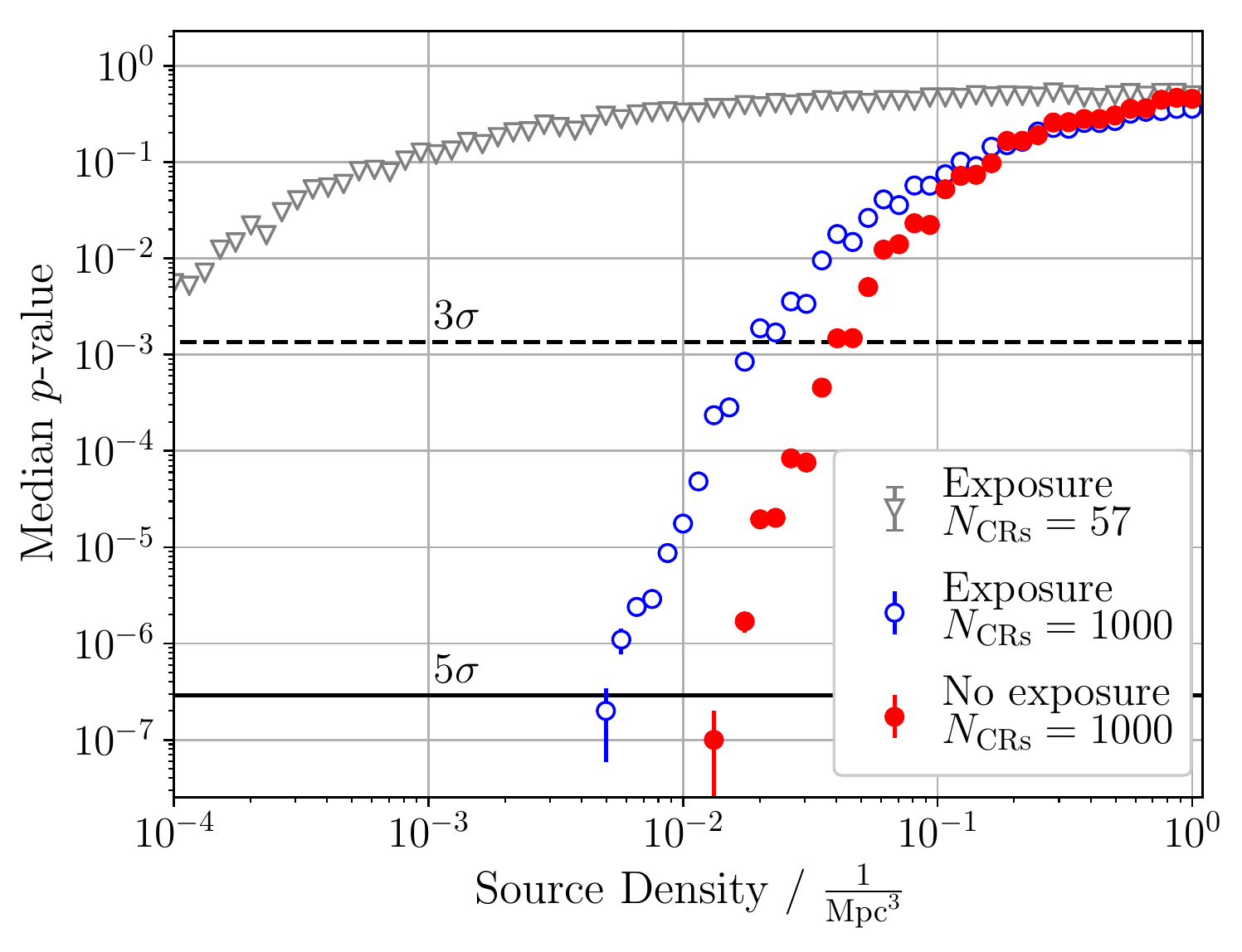}
\caption{Probability of the cosmic-ray arrival directions with multiple signal patterns to originate from isotropic scenarios as a function of the source density. The signal patterns were simulated using the general astrophysical scenario with the source density as the only parameter (section~\ref{sec:astrophysical-scenario}). }
\label{fig:density}
\end{center}
\end{figure}

In Fig.~\ref{fig:density} we show the median $p$-value, determined from the $1000$ signal scenarios, as a function of the source density. Under the conditions of this general astrophysical scenario, a set of 1000 arriving cosmic rays with signal patterns can be identified with $3$~standard deviations at a source density of $4.0\cdot10^{-2}$~Mpc$^{-3}$. The limit of $5$ standard deviations is reached at a source density of $1.5\cdot10^{-2}$~Mpc$^{-3}$.

If data of an observatory which can observe only a part of the sphere are used, the sensitivity decreases accordingly. Here, we use the geometric exposure of the AGASA Observatory by way of example while keeping $N_\textrm{CRs}$ at $1000$. The blue symbols in Fig.~\ref{fig:density} denote the corresponding reduced exclusion limits where the $5\sigma$ confidence limit is reached for $5\cdot10^{-3}$~Mpc$^{-3}$.

Additionally, we explore the sensitivity of the network on a dataset with only $N_\mathrm{CRs}=57$ cosmic rays following the AGASA exposure. The corresponding $p$-values are shown in gray in Fig.~\ref{fig:density}. One can see that the sensitivity is considerably reduced as expected. The $3\sigma$ confidence limit is only reached for a source density of $6\cdot10^{-5}$~Mpc$^{-3}$ and the 5$\sigma$ at $8\cdot10^{-6}$~Mpc$^{-3}$.

\section{Source density limit from AGASA data}

In this section we apply the dynamic graph convolutional network to the $57$ ultra-high energy events of the AGASA Observatory (cf. section~\ref{sec:AGASA-data}). To begin with we train the network using simulated datasets with $57$ cosmic rays within the exposure of the observatory. For the training we use the generalized astrophysical scenario (cf. section~\ref{sec:astrophysical-scenario}) with source density as the only free parameter where we determined $\rho_\textrm{S}=10^{-4}$~Mpc$^{-3}$ to be a good choice based on a scan on simulations. To account for the difference between the energy spectrum of the AGASA data and the simulated one, we replace the cosmic-ray energies before applying the deflection in the simulation: for this, 57 random energies are sampled from a power law spectrum with $\gamma=-3.4$, which results in a median energy close to the median of the AGASA measurements. The energies are replaced such that the order of the energy values remains the same.

For the measured $57$ events, we determine the $p$-value as presented in section~\ref{sec:sensitivity}. For that we generate isotropic sets of $57$ cosmic rays by randomizing the arrival directions of the measured events. This procedure again ensures that the energy distribution of the cosmic rays remains unchanged. 

The network response $x_\textrm{sig}$ from the AGASA dataset lies fairly within the isotropic expectation and is exceeded by $p=65\%$ of the isotropic simulations. Thus, in the relatively few events of the AGASA Observatory the network finds no signs of signal patterns in the arrival directions. 
Therefore, we calculate a lower limit for the source density for the generalized astrophysical scenario. To this end, we invert the question and ask what minimum source density $\rho_\textrm{S}$ the generalized astrophysical scenario needs to have in order to satisfy the $p$-value of the data. For this purpose, we investigate the distribution of the $p$-values for $1000$ simulated data sets at each source density $\rho_\textrm{S}$ of the generalized astrophysical scenario. 

We obtain a lower limit of $3\cdot10^{-4}$~Mpc$^{-3}$ for the density of cosmic-ray sources at $95\%$ confidence level. 

According to Fig.~\ref{fig:density}, we expect that a larger number of cosmic rays, such as those measured by modern observatories, will significantly improve the determination of the source density. In addition, signal patterns could be detected by the higher data statistics.

\section{Conclusion}
In this work we investigated graph networks for the identification of source patterns in the cosmic-ray arrival directions and energies. These source patterns originate from cosmic nuclei that are deflected in the coherent magnetic field of our Galaxy and allow cosmic-ray sources to be identified.

Graph convolutional networks are particularly well suited to investigate cosmic rays with their individual arrival directions without digitizing the directions on a Cartesian or spherical grid. The convolutional operation is defined by the particles' nearest neighbors, whose properties are characterized in a $K$-dimensional space. 

When classifying a data set as either isotropic background or containing a signal pattern, our investigations show that the particles of a signal pattern are bundled in the $K$-dimensional property space separately from the isotropically arriving particles. 

The network thus identifies data sets with a signal pattern of $18$ helium nuclei in $982$ isotropically distributed particles with a confidence of $5$ standard deviations against isotropic data sets. Additionally, in median the network correctly identifies all $18$ helium nuclei from the signal pattern and incorrectly identifies $12$ particles that actually belong to the isotropic background.

For a generalized astrophysical scenario, which depends only on the source density and generates multiple source patterns distributed over the sphere, dynamic graph convolutional networks prove to be advantageous. In this case, the neighborhoods of the particles in deeper network layers can be optimized for the convolutional operation in such a way that the properties of signal patterns can be studied together despite very different spatial arrival directions.

As a measure for the sensitivity of such networks, we expect that the existence of signal patterns can be proven with $5$ standard deviations for $1000$ cosmic rays at an upper limit for the source density of $1.5\cdot10^{-2}$~Mpc$^{-3}$.

As an example for data measured by an observatory we analyzed the public data of the AGASA experiment with energies above $40$~EeV. A signal pattern could not be detected here with 57 cosmic rays only. However, from these data we give a lower limit on the source density of the generalized astrophysical scenario of $\rho_\textrm{S}^{95}=3\cdot10^{-4}$~Mpc$^{-3}$ at $95\%$ confidence level. 

We also show that we could greatly improve such an exclusion limit for the source density when using data from current observatories with more than $1000$ cosmic rays. Furthermore, patterns may be found and thus sources of ultra-high energy cosmic rays identified.

\section*{Acknowledgments}

We wish to thank Yannik Rath for his valuable comments on the manuscript.
This work is supported by the Ministry of Innovation, Science and Research of the State of North Rhine-Westphalia, and by the Federal Ministry of Education and Research (BMBF).

\bibliography{GraphNetwork}

\end{document}